\documentclass[twocolumn]{aastex61}
\usepackage{graphics}
\usepackage[caption=false]{subfig}
\usepackage{amsmath}
\usepackage{hyperref}
\usepackage{amssymb}
\usepackage{xcolor}
\usepackage{epsfig}
\usepackage{latexsym}
\usepackage{tensor}
\usepackage{float}
\usepackage{multirow}
\usepackage{xspace}
\usepackage{comment}
\usepackage{acronym}

\newcommand{\ee}[1]{\ensuremath{\!\times\!10^{#1}}}

\newcommand{\grbrate}{{{\mathcal R}_{\mathrm{grb}}}}
\newcommand{\cbcrate}{{{\mathcal R}}}
\newcommand{\diff}{{\mathrm d}}

\newcommand{\dhor}{\ensuremath{{\mathcal D}_{\mathrm{hor}}}}
\newcommand{\dinsp}{\ensuremath{{\mathcal D}_{\mathrm{insp}}}}
\newcommand{\latin}[1]{\textit{#1}}
\newcommand{\mpc}{\mathrm{Mpc}}
\newcommand{\yr}{\mathrm{yr}}

\newacro{aLIGO}[aLIGO]{Advanced LIGO}
\newacro{GRB}[GRB]{gamma-ray burst}
\newacro{SGRB}[SGRB]{short gamma-ray burst}
\newacro{SGR}[SGR]{soft gamma-ray repeater}
\newacro{GW}[GW]{gravitational wave}
\newacro{NS}[NS]{neutron star}
\newacro{BH}[BH]{black hole}
\newacro{SNR}[SNR]{signal-to-noise ratio}
\newacro{MWEG}[MWEG]{Milky Way Equivalent Galaxy}
\newacro{KDE}[KDE]{kernel density estimation}
\newacro{MAP}[MAP]{Maximum \latin{a posteriori}}

\newcommand{\BNS}{\ac{NS}--\ac{NS}\xspace}
\newcommand{\BNSmm}{{\small\ac{NS}-\ac{NS}}} 
\newcommand{\NSBH}{\ac{NS}--\ac{BH}\xspace}
\newcommand{\JOINT}{\ac{GW}--\ac{SGRB}\xspace}

\def\imbh#1{intermediate mass black hole#1(IMBH#1)\gdef\imbh{IMBH}}
\def\smbh#1{supermassive black hole#1(SMBH#1)\gdef\smbh{SMBH}}
\def\bbh#1{binary black hole#1 (BBH#1)\gdef\bbh{BBH}}
\def\bh#1{black hole#1 (BH#1)\gdef\bh{BH}}
\def\sn#1{core-collapse supernova#1 (CCSN#1)\gdef\sn{CCSN}}
\def\pnw#1{post-Newtonian#1 (PN#1)\gdef\pnw{PN}}
\def\eos#1{equation of state#1 (EOS#1)\gdef\eos{EOS}}
\def\electro#1{electromagnetic#1 (EM#1)\gdef\electro{EM}}
\def\amr#1{adaptive mesh refinement#1 (AMR#1)\gdef\amr{AMR}}
\def\isco#1{innermost stable circular orbit#1 (ISCO#1)\gdef\isco{ISCO}}
\def\cwb#1{Coherent WaveBurst#1 (CWB#1)\gdef\cwb{CWB}}

\definecolor{dwnote}{HTML}{E24A33}

\def\aap{Astronomy and Astrophysics}

\definecolor{dgreen}{rgb}{0.3, 0.73, 0.09}
\usepackage{soul} 
\usepackage{ulem} 

\begin{document}

\title{Constraints On Short, Hard Gamma-Ray Burst Beaming Angles From
Gravitational Wave Observations}

\author[0000-0003-3772-198X]{D. Williams}
\affiliation{SUPA, University of Glasgow, Glasgow G12 8QQ, United Kingdom}
\author[0000-0003-3243-1393]{J. A. Clark}
\affiliation{Center for Relativistic Astrophysics and School of Physics, Georgia Institute of Technology, Atlanta, GA 30332}
\author[0000-0002-7627-8688]{A. R. Williamson}
\affiliation{Department of Astrophysics/IMAPP, Radboud University Nijmegen, P.O. Box 9010, 6500 GL Nijmegen, Netherlands}
\author[0000-0002-1977-0019]{I. S. Heng}
\affiliation{SUPA, University of Glasgow, Glasgow G12 8QQ, United Kingdom}

\correspondingauthor{Daniel Williams}
\email{d.williams.2@research.gla.ac.uk}


\begin{abstract}
The first detection of a binary neutron star merger, GW170817, and an associated
short gamma-ray burst confirmed that neutron star mergers are responsible for at
least some of these bursts. The prompt gamma ray emission from these events is
thought to be highly relativistically beamed. We present a method for inferring
limits on the extent of this beaming by comparing the number of short gamma-ray
bursts observed electromagnetically to the number of neutron star binary mergers
detected in gravitational waves.
We demonstrate that an observing run comparable to the expected Advanced LIGO
2016--2017 run would be capable of placing limits on the beaming angle of
approximately $\theta \in (2.88^\circ,14.15^\circ)$, given one binary neutron
star detection.
We anticipate that after a year of observations with Advanced LIGO at design
sensitivity in 2020 these constraints would improve to
$\theta \in (8.10^\circ,14.95^\circ)$. 
\end{abstract}


\section{Introduction}

Gamma-ray bursts (GRBs)\acused{GRB} are extremely energetic cosmological events
observed approximately once per day. There appear to be at least two separate
populations of \acp{GRB}, divided roughly according to their duration and
spectral hardness~\cite{Kouveliotou:1993yx}, although with significant overlap
obscuring any clear distinction between
populations~\cite{Zhang:2009uf,Bromberg:2012gp}.
Those with long durations ($\gtrsim 2\,\mathrm{s}$) and softer spectra are associated
with core collapse
supernovae~\cite{Galama:1998ea,MacFadyen:1998vz,Woosley:2006fn}. Short, hard
\acp{GRB} (SGRBs)\acused{SGRB} were long suspected of being the signatures of
compact binary coalescences involving at least one
\ac{NS}~\cite{Blinnikov1984,Eichler:1989ve,Paczynski:1991aq,Narayan:1992iy,Lee:2007js}.
Both \BNS and \NSBH progenitors are possible, with the requirement that a post
merger torus of material accretes onto a compact central
object~\cite{Blandford:1977ds,Rosswog:2002rt,Giacomazzo:2012zt}.

The first observation of a \BNS coalescence event,
GW170817~\cite{TheLIGOScientific:2017qsa}, its association with
GRB~170817A~\cite{Monitor:2017mdv,Goldstein:2017mmi,Savchenko:2017ffs}
and, later, multi-wavelength electromagnetic emission, including a
kilonova~\cite{2017ApJ...848L..12A}, confirmed that compact binary mergers are
the engines of at least some \acp{SGRB}. The \ac{GW} observation placed only
weak constraints on the viewing angle due to a degeneracy between distance and
inclination of the binary to the line of sight~\cite{TheLIGOScientific:2017qsa}.
However, GRB~170817A was not typical of \acp{SGRB}, being around $10^{4}$ times
less energetic~\cite{Goldstein:2017mmi}. This, in addition to other aspects of
the electromagnetic emission, has been widely interpreted as indicating that
GRB~170817A was not viewed from within the cone of a canonical jet with a
top-hat profile (see
e.g.~\cite{Fong:2017ekk,Kasliwal:2017ngb,Gottlieb:2017pju,Haggard:2017qne}).

A population of \JOINT observations could also allow us to measure the fraction
of \acp{SGRB} associated with each progenitor type, and associated redshifts
will enable a relatively systematics-free measurement of the Hubble parameter
at low redshift, which would provide constraints on cosmological
models~\cite{Schutz:1986gp,Nissanke:2009kt,Chen:2012qh,Abbott:2017xzu}.

In this work we consider a population of binary merger sources, with and without
\ac{SGRB} counterparts, assuming that the vast majority of these counterparts
would be viewed from within the cone of a standard jet. This is motivated by the
fact that most mergers would be expected to occur at distances much greater than
GW170817 and that weak, off-axis gamma ray emission would in all likelihood go
undetected. With such a population we can constrain the average opening
angle~\cite{Chen:2012qh,Clark:2014jpa,Abbott:2016ymx}.

We investigate what statements can \emph{currently} be made on the beaming angle
itself using the bounds placed on the binary merger rate $\cbcrate$ from
all-sky, all-time \ac{GW} searches and explore the potential for direct
inference of \ac{SGRB} beaming angles in the advanced detector era.
We first discuss the relationship between \acp{SGRB} and compact binary
coalescences. In particular, we will focus on \BNS inspirals as the progenitors
of \acp{SGRB}. We then present our method for robustly inferring the jet opening
angles using only \ac{GW} observations. We demonstrate our method assuming the
nominal number of \ac{GW} signals observed from \BNS inspirals expected for
\ac{aLIGO} and Advanced Virgo in planned observing scenarios, as defined
in~\cite{Aasi:2013wya}. Finally, we conclude with a discussion on the
implications of our work as well as possible avenues for further extension of
the work presented here.

\section{Short gamma-ray bursts and compact binary coalescences}
\label{sec:sgrbs}
At their design sensitivities, the current generation of advanced
\ac{GW} detectors could observe \BNS mergers out to distances of
${\sim}400\,\mpc$ at a rate of $0.1\text{--}200\,$yr$^{-1}$~\cite{Aasi:2013wya}.
It is worth noting that at galactic or near-galactic distances, \ac{SGR}
hyperflares can resemble \acp{SGRB}. These \ac{SGR} hyperflares are the likely
explanations for GRB~070201 and GRB~051103, since compact binary coalescences at
the distance of their probable host galaxies were excluded with
greater than $90\%$ confidence \cite{Abbott:2007rh,Abadie:2012bz}.

Given the link between \acp{SGRB} and compact binary coalescences, it is
interesting to ask whether the \ac{SGRB} beaming angle can be inferred from
\ac{GW} observations. As discussed in~\cite{Clark:2014jpa}, a comparison of
the populations of observed \acp{SGRB} and \BNS mergers may be the most
promising avenue for this. Motivated by the study in~\cite{Chen:2012qh},
we note that if the \ac{SGRB} population posseses a distribution of beaming
angles then the \emph{observed} rate of \acp{SGRB} is related to the rate of
\BNS coalescences $\cbcrate$ via,
\begin{equation}\label{eq:rate2angle}
    \grbrate = \epsilon\cbcrate \left \langle 1-\cos \theta \right \rangle,
\end{equation}
where angled brackets $\langle \rangle$ indicate the population mean
and $\epsilon$ is the probability that a binary coalescence results in
an observed \ac{SGRB}.  In this work, we assume an illustrative
$\grbrate=10$\,Gpc$^{-3}$\,yr$^{-1}$~\cite{Nakar:2007yr,Dietz:2010eh}
and we shall refer to $\epsilon$ as the \ac{SGRB}
\emph{efficiency}. The method we present, however, is amenable to
using alternative values for $\grbrate$, or indeed to being extended to
sampling values from a prior distribution on the \ac{SGRB} rate. Generally,
the efficiency with which \BNS mergers produce \acp{SGRB} is unknown
but will depend on a variety of progenitor physics.  In particular, a
significant fraction of \NSBH systems may be incapable of powering an
\ac{SGRB}~\cite{Pannarale:2014rea}.  Combining this knowledge with
measurements of the binary parameters of of a population of \JOINT
observations could be used to constrain $\epsilon$.  In this work, we
will make no attempt to characterise $\epsilon$ and we simply aim to
provide a framework which allows one to incorporate various levels of
assumptions (or ignorance) regarding its value.

\begin{figure}
\centering
\includegraphics[width=\linewidth]{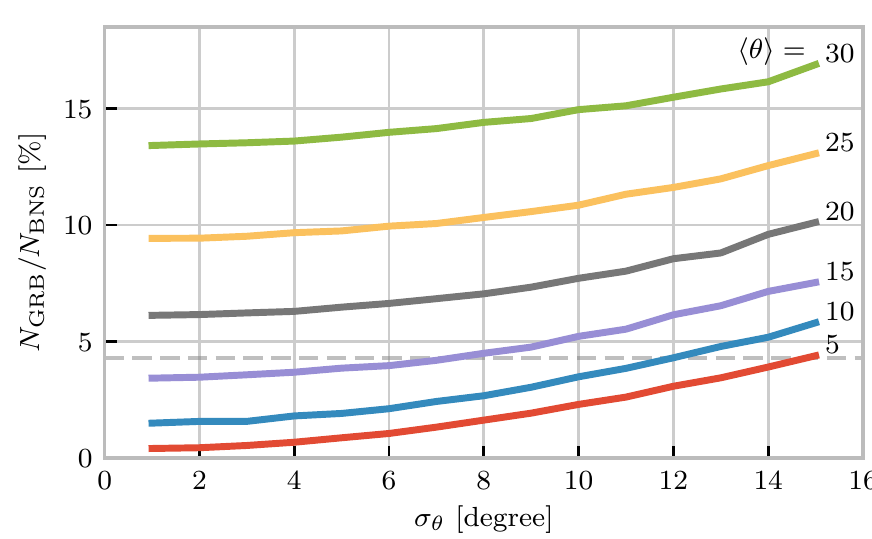}
\caption{\label{fig:thetapopulation} Expected relative numbers of
  observed GRBs and binary coalescences for different distributions on
  the GRB beaming angle.  Lines in the figure correspond to jet angle
  population means, while the $x$-axis shows the width of the
  distribution.  All distributions are Gaussian, truncated at
  $(0, 90]$ degrees.\label{fig:thetapop}}
\end{figure}

If the \ac{SGRB} population has a distribution of beaming angles, as would seem
likely from \electro{} observations~\cite{Fong:2015oha}, characterising the
relative rates of \ac{SGRB} and \BNS coalescence will inform us as to the mean
of that population, $\langle \theta \rangle$.  To explore this point further we
construct a simple Monte Carlo simulation to study the effect on the relative
rates of \acp{SGRB} and \BNS mergers. We arrange the following toy problem:
\begin{enumerate}
    \item Set the number of `observed' \acp{SGRB} to zero: $N_{\mathrm{GRB}}=0$.
    \item Draw $N_{\mathrm{\BNSmm}}$ values of orbital inclination $\iota$ from a distribution which is uniform in $\cos \iota$ in the range $[0,1]$.
    \item For each value of $\iota$, draw a value for the beaming angle $\theta$, from some distribution with finite width and limited to the range $(0,90]^{\circ}$.
    \item If $\iota<\theta$ then this combination of orbital inclination and beaming angle would result in an observable \ac{SGRB}, so increment $N_{\mathrm{GRB}}$.
\end{enumerate}
Such a simulation allows us to study the ratio of the number of
observed \ac{SGRB} to the total number of \BNS mergers
$N_{\mathrm{GRB}}/N_{\mathrm{\BNSmm}}$.  Since it is the comparison of
the rates of these events that informs our inference on $\theta$,
studying the ratio $N_{\mathrm{GRB}}/N_{\mathrm{\BNSmm}}$ provides some
intuition as to the effect and features of various $\theta$
distributions.  Figure~\ref{fig:thetapop} plots this ratio as a
function of various truncated normal distributions to demonstrate the
effect of shifting the mean and scaling the width of the distribution.
Points along the $x$-axis correspond to different choices of the
distribution width $\sigma_{\theta}$, and the separate curves
correspond to different choices of the distribution mean
$\langle \theta \rangle$.  Let us denote this truncated normal
distribution ${\mathcal N}(\langle \theta \rangle, \sigma_{\theta})$.
We stress here that such $\theta$ distributions are \emph{not}
intended to represent the true distribution; they are merely intended
to easily demonstrate the qualitative effects of different $\theta$
distributions on the ratio $N_{\mathrm{GRB}}/N_{\mathrm{\BNSmm}}$.

Figure~\ref{fig:thetapop} reveals that a population of \ac{SGRB} beaming angles with a large mean but narrow width is, on the basis of rate measurements, indistinguishable from a population of \ac{SGRB} beaming angles with a small mean and large width.
For example, for the ${\mathcal N}(15,9)$ and ${\mathcal N}(10,13)$ beaming angle populations, the ratios of $N_{\mathrm{GRB}}/N_{\mathrm{\BNSmm}}$ are almost equal ($\sim 4.8\%$).
Thus, a sufficiently wide spread of \ac{SGRB} beaming angles will yield relatively high rates for \BNS and \acp{SGRB} that could lead to an overestimate of the mean beaming angle.
The population-based constraints on $\theta$ must, therefore, be regarded as upper bounds on the mean of a distribution of beaming angles.
Having said this, for a given mean value $\langle \theta \rangle$, the ratio is rather insensitive to the width.

\section{From Rates To Beaming Angles}

In this section, we discuss our approach to estimating the \ac{SGRB} beaming angle based on the binary neutron star inspiral rate, estimated through a number of \ac{GW} observations of \BNS coalescence.
We demonstrate the approach by considering plausible detection scenarios for \ac{aLIGO}~\cite{Aasi:2013wya}.
Our ultimate goal is to develop a generic approach that folds in uncertainties in the \BNS merger rate and our ignorance about the probability with which such mergers actually result in \acp{SGRB}.
An overview of the general method is as follows:

\begin{enumerate}
    \item Estimate the posterior probability distribution on the \BNS merger rate
    in the local universe from a number of observed gravitational wave signals
    and our knowledge of the sensitivity of the detectors.  We construct a joint
    posterior distribution on the \BNS rate and the (unknown) probability
    $\epsilon$ that a given merger results in a \ac{SGRB}.
\item Use equation~\ref{eq:rate2angle}, which relates the \BNS merger and
    \ac{SGRB} rates via the geometry of the beaming angle, to transform the rate
    posterior probability to a posterior probability on the mean \ac{SGRB}
    beaming angle.
\item Marginalize over $\epsilon$. We choose to consider $\epsilon$ a nuisance
    parameter because, to date, there is no accurate estimate of this parameter
    and it is not the main focus of our analysis. 
\end{enumerate}

\subsection{Constructing The Rate Posterior}
\label{sec:rate_posterior}

Our goal is to infer the posterior probability distribution for the mean
\ac{SGRB} beaming angle $\theta$ from \ac{GW} constraints on the rate of \BNS
coalescence $\cbcrate$.  The core ingredient to the analysis is the posterior
probability distribution on the coalescence rate $p(\cbcrate|D,I)$, where $D$
represents some \ac{GW} observation and $I$ denotes other unenumerated prior
information.  We will first demonstrate how $p(\cbcrate|D,I)$ may be constructed
for a few projected observing scenarios from~\cite{Aasi:2013wya}.  Later, in
section~\ref{sec:beaming_limits}, we will extend the analysis to place
limits on $\theta$ based upon the lack of detection during O1. Previously, a
comparison of rates was used to place a lower limit on the beaming angle
in~\cite{Abbott:2016ymx}.

To form the posterior on the coalescence rate, we begin by constructing the
posterior on the \emph{signal} rate.  Note that these are not identical since
only those \BNS mergers which occur within a certain range yield a detectable
signal.  \ac{GW} data analysis pipelines (e.g. {\tt
FINDCHIRP}~\cite{2012PhRvD..85l2006A}, {\tt
PyCBC}~\cite{Canton:2014ena,Usman:2015kfa,alex_nitz_2017_844934}) identify
discrete `candidate events' which are characterised by network \acp{SNR},
$\rho_c$, which, for the case of \BNS searches, indicate the similarity between
the detector data and a set of template \BNS coalescence waveforms.  The
measured rate $r$ of these events consists of two components: a population of
true \ac{GW} signals, $s$; and a background rate, $b$, due to noise fluctuations
due to instrumental and environmental disturbances.
\begin{equation}
r = s + b
\begin{cases}
s = \text{signal rate} \\
b = \text{background rate}.
\end{cases}
\end{equation}
Typically for an all-sky, all-time analysis, like that described
in~\cite{Usman:2015kfa}, the significance of a candidate event is
empirically measured against `background' data representative of the
detector noise, which naturally varies from candidate to candidate.  A
detection requires this significance to be above some pre-determined
threshold (e.g. $5\sigma$ for GW150914 and
GW151226~\cite{Abbott:2016blz,Abbott:2016nmj}).  We follow the method
in~\cite{Aasi:2013wya}, which defines a detection as a candidate with
$\rho_c \geq 12$, corresponding approximately to
$b=10^{-2}$\,yr$^{-1}$.  Since the background rate $b$ is known, we
are just left with the problem of inferring the signal rate $s$.
Assuming a uniform prior on $s$ and a Poisson process underlying the
events, it may be shown (e.g.,~\cite{2010blda.book.....G}) that the
posterior for the signal rate, given a known background rate $b$ and
$n$ events observed over a time period $T$ is,
\begin{equation}
p(s|n,b,I) = C \frac{ T\left[(s+b)T\right]^n e^{-(s+b)T}}{n!},
\end{equation}
where,
\begin{eqnarray}
C^{-1} & = &\frac{e^{-bT}}{n!} \int_0^{\infty}\diff(sT)(s+b)^n T^n e^{-sT}\\
& = & \sum_{i=0}^n \frac{ (bT)^i e^{-bT}}{i!}.
\end{eqnarray}
Finally, we can transform the posterior on the \emph{signal} rate to
the underlying \emph{coalescence} rate via our knowledge of the
sensitivity of the \ac{GW} analysis.  In particular, the signal
detection rate is simply the product of the intrinsic coalescence rate
$\cbcrate$ and the number of \BNS mergers which would result in a
\ac{GW} signal with $\rho_c\geq12$.  Expressing the binary coalescence
rate in terms of the number of mergers per \ac{MWEG}, per year then we
require the number of galaxies $N_{\mathrm{G}}$ which may be probed by
the \ac{GW} analysis.  At large distances, this is well approximated
by~\cite{rates_paper},
\begin{equation}
    N_G = \frac{4}{3} \pi \left( \frac{\dhor}{\mpc}
\right)^3 (2.26)^{-3} (0.0116),
\end{equation}
where $\dhor$ is the horizon distance (defined as the distance at which an
optimally-oriented \BNS merger yields $\rho_c\geq12$), the factor of 2.26
results from averaging over sky-locations and orientations, and
$1.16\times 10^{-2}$\,Mpc$^{-3}$ is the extrapolated density of \ac{MWEG} in
space.

Finally, the posterior on the binary coalescence rate $\cbcrate$ is obtained from a trivial transformation of the posterior on the signal rate $s$,
\begin{eqnarray}
    p(\cbcrate|n,T,b,\dhor) & = & p(s|n,T,b) \left|\frac{\diff s}{\diff \cbcrate}\right| \\
                                   & = & N_G(\dhor)p(s|n,T,b).
\end{eqnarray}
We see that in this approach, the rate posterior depends only on the
number of signal detections $n$, the observation time $T$, the
background rate $b$, and the horizon distance of the search $\dhor$.
It is precisely these quantities that comprise the detection scenarios
outlined in~\cite{Aasi:2013wya}.  Before constructing expected rate
posteriors, we outline the transformation from rate to beaming angle.

\subsection{Constructing the beaming angle posterior}
Inferences of the \ac{SGRB} beaming angle are made from the posterior
probability density on the beaming angle $p(\theta|D,I)$ where, as
usual, $D$ indicates some set of observations and $I$ unenumerated
prior knowledge.  Our goal is to transform the measured posterior
probability density on the rate $\cbcrate$ to a posterior on the
beaming angle.
First, note that we can express the joint distribution
$p(\theta, \epsilon|D,I)$ as a Jacobian transformation of the joint
distribution $p(\cbcrate, \epsilon|D,I)$:
\begin{equation}
p(\theta,\epsilon) = p(\cbcrate,\epsilon)
\left\lvert\left\lvert
\frac{\partial(\cbcrate,\epsilon)}{\partial(\theta,\epsilon)}
\right\rvert\right\rvert,
\end{equation}
where we have dropped conditioning statements for notational convenience.
The Jacobian determinant can be  computed from equation~\ref{eq:rate2angle}.
It is then straightforward to marginalize over $\epsilon$ to yield the posterior on $\theta$ itself:
\begin{eqnarray}
    \label{eq:beam_posterior}
    p(\theta) & = & \int_{\epsilon} p(\theta,\epsilon)~\diff \epsilon\\
              & = & \int_{\epsilon} p(\cbcrate,\epsilon)
    \left\lvert\left\lvert
    \frac{\partial(\cbcrate,\epsilon)}{\partial(\theta,\epsilon)}
    \right\rvert\right\rvert~\diff \epsilon \\
              & = & \frac{2\grbrate \sin
\theta~p(\cbcrate)}{(\cos\theta-1)^2}\int_{\epsilon}
\frac{p(\epsilon)}{\epsilon} ~\diff \epsilon,
\end{eqnarray}
where we have assumed $\epsilon$ and $\cbcrate$ are logically independent such that,
\begin{equation}
p(\epsilon,\cbcrate) = p(\epsilon|\cbcrate)p(\cbcrate) = p(\epsilon)p(\cbcrate).
\end{equation}
It is important to note that the entire procedure of deriving the jet
angle posterior is completely independent of the approach used to
derive the rate posterior.  In the preceding section we adopted a
straightforward Bayesian analysis of a Poisson rate which is amenable
to a simple application of plausible future detection scenarios; there
is no inherent requirement to use that method to derive the rate
posterior.

Given the posterior on the rate, $p(\cbcrate)$, the final ingredient
in this approach is the specification of some prior distribution for
$\epsilon$. Given the lack of information on the value and distribution
of $\epsilon$, we choose three plausible priors and study their effects
on our beaming angle inference. Our choice of priors are:
\begin{description}
\item [Delta-function] $p(\epsilon) = \delta(\epsilon=0.5)$;
        the probability that \BNS mergers yield \acp{SGRB} is known to be 50\%
        exactly.

\item [Uniform] $p(\epsilon)=U(0,1)$;
        the probability that \BNS mergers yield \acp{SGRB} may lie anywhere
    $\epsilon \in (0,1]$ with equal support in that range. 

    \item [Jeffreys] $p(\epsilon)=\beta(\frac{1}{2},\frac{1}{2})$; treating the
        outcome of a \BNS merger as a Bernoulli trial in which a \ac{SGRB}
        constitutes `success' and $\epsilon$ is the probability of that success,
        the least informative prior, as derived from the square root of the
        determinant of the Fisher information for the Bernoulli distribution, is
        a $\beta$-distribution with shape parameters $\alpha=\beta=\frac{1}{2}$.
\end{description}

\section{Prospects For Beaming Angle Constraints With Advanced LIGO}
We now demonstrate the derivation of the rate posterior $p(\cbcrate)$
and the subsequent transformation to the beaming angle posterior
$p(\theta)$.  We consider four \ac{GW} observation scenarios with
\ac{aLIGO} based on the work in~\cite{Aasi:2013wya}.  An observing
scenario essentially consists of an epoch of \ac{aLIGO} operation,
which defines an expected search sensitivity (i.e., \BNS{} horizon
distance $\dhor$) and observation time $T$; as well as an assumption
on the rate of \BNS{} coalescence in the local universe $\cbcrate$.
Each observing scenario ultimately results in an expectation for the
number of observed \acp{GW} from \BNS coalescences.  For this study,
we assume the `realistic rate' for $\cbcrate$ as described
in~\cite{rates_paper}.

Our first goal is to establish the expected number of detections in
each scenario.  Given the observation time and horizon distance of the
observation epoch we first compute the 4-volume accessible to the
analysis,
\begin{equation}
    \label{eq:search_volume}
    V_{\mathrm{search}} = \frac{4}{3}\pi \left(\frac{\dhor}{2.26}\right)^3 \times \gamma T,
\end{equation}
where the factor 2.26 arises from averaging over source sky location
and orientation, $T$ is the observation time and $\gamma$ is the
\emph{duty cycle} for the science run.  Following~\cite{Aasi:2013wya},
we take $\gamma=0.5$.  For comparison, during the first observing run
of \ac{aLIGO}, the two interferometers observed in coincidence
achieving $\gamma_{\mathrm{coinc}} = 0.41$.  Where there is a range in
the horizon distances quoted in~\cite{Aasi:2013wya} to account for
uncertainty in the sensitivity of the early configuration of the
detectors, we use the arithmetic mean of the lower and upper bounds
when computing the search volume.  Table~\ref{tab:scenarios} lists
the details of each observing scenario.
\begin{table}
\centering
\begin{tabular}{lccccc}
  \toprule
  Epoch &  $T$ & \dinsp & $V_{\mathrm{search}}$ & Est. \BNS \\
        &   [yr] & [Mpc] & [$\ee{6} \mpc³\,\yr^{-1}$] & Detections \\
  \colrule
  2015--2016 & 0.25 & 40--80   & 0.05--0.4 & 0.0005--4 \\
  2016--2017 & 0.5 & 80--120 & 0.6--2.0 & 0.006-20\\
  2018--2019 & 0.75 & 120--170 & 3--10 & 0.04--100\\
  2020+      & 1    & 200 & 20 & 0.2--200 \\
  2024+      & 1    & 200 & 40 & 0.4--400 \\
  \botrule
\end{tabular}
\caption{Advanced detector era observing scenarios considered in this
  work.  $T$ is the expected duration of the science run and $\dinsp$
  is the \BNS inspiral distance for the sensitivity expected to be
  achieved at the given epoch, which is equal to $\dhor / 2.26$.
  $V_{\mathrm{search}}$ is the sensitive volume of the search, defined
  by equation~\ref{eq:search_volume}; the final column contains the
  estimated range of the number of \ac{GW} detections.  Note that the
  quoted search volume accounts for a network duty cycle of
  $\sim 80\%$ per detector.  These scenarios are derived from those detailed
  in~\cite{Aasi:2013wya}. While the 2020+ and 2024+ scenarios appear
  identical in terms of the sensitivity of the detectors, the 2024+
  scenario includes a third advanced LIGO detector in India. This
  expansion of the network is expected to lead to an increase in the
  network duty cycle, and an increase in the area of the sky which the network is sensitive to, resulting in a greater volume being searched per
  year.
  \label{tab:scenarios}}
\end{table}

\subsection{Posterior Results}
\begin{figure}
\centering
{\includegraphics[width=\linewidth]{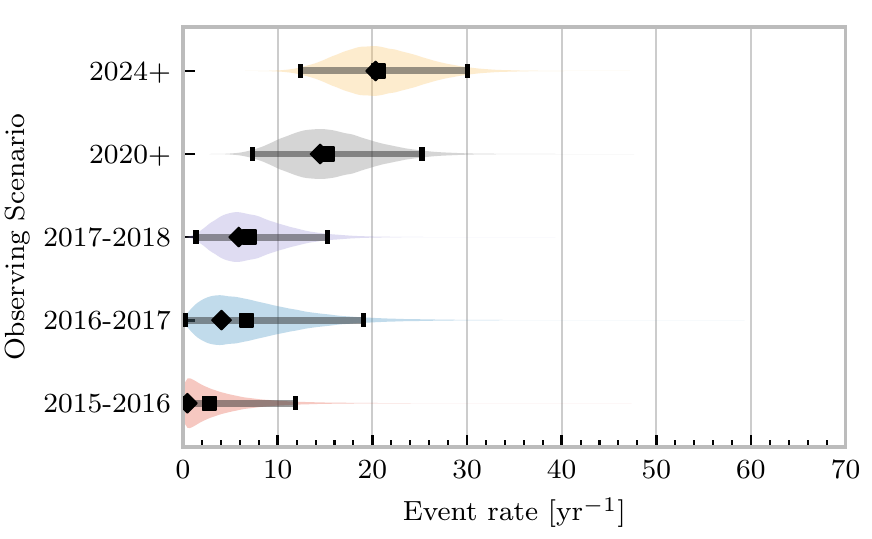}}
\caption{Posterior probability distribution for the rate of \BNS
    coalescence assuming the scenarios in table \ref{tab:scenarios}.
    The 95\% credible interval is represented with a horizontal line through
    the centre of the plot, with vertical lines delineating the lower and upper limits; the median is represented by a square marker, and the
    maximum \latin{a posteriori} (\ac{MAP}) value is denoted by a diamond. A
    summary of these values is given in table \ref{tab:rateposteriors}.
    \label{fig:aligorate} }
\end{figure}

Figure~\ref{fig:aligorate} shows the \BNS rate posteriors resulting
from the observations in the scenarios in table~\ref{tab:scenarios}
generated using the procedure described in
section~\ref{sec:rate_posterior}. Where a range of potential inspiral
distances is given for a scenario we choose the median value, so for
the 2015--2016 scenario we take \dinsp{} to be $60\,\mpc$, for
example. Likewise we choose an illustrative value of $n$, the number
of expected \ac{GW} detections, from each range; these are listed in
table \ref{tab:rateposteriors}.

We now use these posteriors together with the prior distributions
described in section~\ref{sec:rate_posterior} and the observed rate of
\acp{SGRB} (as described in section~\ref{sec:sgrbs}, we use
$\grbrate=10$\,Gpc$^{-3}$yr$^{-1}$~\cite{Nakar:2007yr,Dietz:2010eh})
to derive the corresponding beaming angle posteriors.

\begin{table}
\begin{center}
  \begin{tabular}{lrrrrr}
    \toprule
    Scenario &    $n$ & Lower       & MAP             & Median          & Upper\\
             &        & [$\yr^{-1}$] & [$\yr^{-1}$]    & [$\yr^{-1}$]    & [$\yr^{-1}$]  \\
    \colrule
    2015--2016 & 0   & 0.00	 & 0.45	 & 2.80	 & 11.98	\\
    2016--2017 & 1   & 0.17	 & 4.07	 & 6.74	 & 19.13	\\
    2017 -- 2018 & 3 & 1.37	 & 5.88	 & 6.99	 & 15.26 \\	
    2020+ & 10 &7.30	 & 14.47	 & 15.25	 & 25.25	\\
    2024+ & 20 & 12.42	 & 20.35	 & 20.65	 & 30.09	\\
    \botrule
\end{tabular}
\end{center}

\caption{Summary of the \BNS rate posteriors for each of the observing
  scenarios which are considered in this work; these posteriors are plotted
  in figure \ref{fig:aligorate}. Here $n$ is the number of \ac{GW} events which were assumed to be observed in each scenario, chosen from the ranges in table \ref{tab:scenarios}.
  \label{tab:rateposteriors}
}
\end{table}

\subsubsection{Validation}
Before we derive beaming angle posteriors corresponding to the
aforementioned observing scenarios, it is useful to establish some
form of validation for our procedure.  This validation is performed by
first selecting values of the beaming angle, the \ac{SGRB} efficiency,
and the rate of \BNS coalescence.  We choose $\theta=10^{\circ}$, and
the `realistic' \BNS rate $\cbcrate = 10^{-6}$\,Mpc$^{-3}$yr$^{-1}$.
We then compute the value of the \ac{SGRB} rate that would correspond
to these parameter choices.  Finally, we simply use this
\emph{artificial} value for $\grbrate$ in
equation~\ref{eq:beam_posterior} when we compute the posterior on the
beaming angle, with the understanding that the resulting posterior
should yield an inference consistent with the `true' value
$\theta=10^{\circ}$.
\begin{figure}
\centering
{\includegraphics[width=\linewidth]{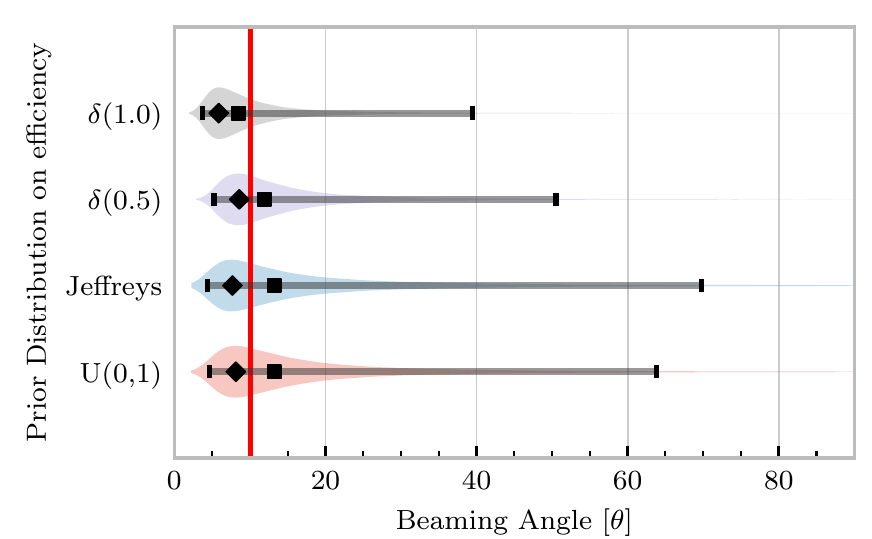}}
\caption{ In order to validate the algorithm an artificial scenario
  was constructed with a known beaming angle by artificially setting a
  GRB rate of $36.7\, \yr^{-1}$ to induce a beaming angle of $\theta \approx 10^{\circ}$.
  The algorithm was then tested with the various priors used in the
  analysis,  using
  the same horizon distance, observing time, and duty cycle as the 2015--2016
  observing scenario. to ensure that the correct beaming angle was inferred. 
  These posteriors are based on the simulated 2015--2016 observing scenario (see
  table~\ref{tab:scenarios}).
  \label{fig:injjetposterio2016}}
\end{figure}

\begin{table}
  \centering
  \begin{tabular}{lrrrr}
    \toprule
    Prior & Lower & MAP & Median & Upper\\
          & [$^\circ$] & [$^\circ$]& [$^\circ$]& [$^\circ$] \\
    \colrule
    $\delta(1.0)$ & 3.68	 & 5.88	 & 8.45	         & 39.44	 \\
    $\delta(0.5)$ & 5.24	 & 8.59	 & 11.89	 & 50.51	 \\
    Jeffreys      & 4.38	 & 7.69	 & 13.23	 & 69.74	 \\
    U(0,1)        & 4.62	 & 8.14	 & 13.23	 & 63.81	 \\
    \botrule
\end{tabular}
\caption{Summary of the beaming angle posteriors from figure
  \ref{fig:injjetposterio2016}, for the 2015--2016 observing scenario,
  with an artificial GRB rate imposed to produce a target beaming
  angle of $\theta = 10^{\circ}$.
  \label{tab:summaryinj2015}}
\end{table}

\begin{figure}
\centering
{\includegraphics[width=\linewidth]{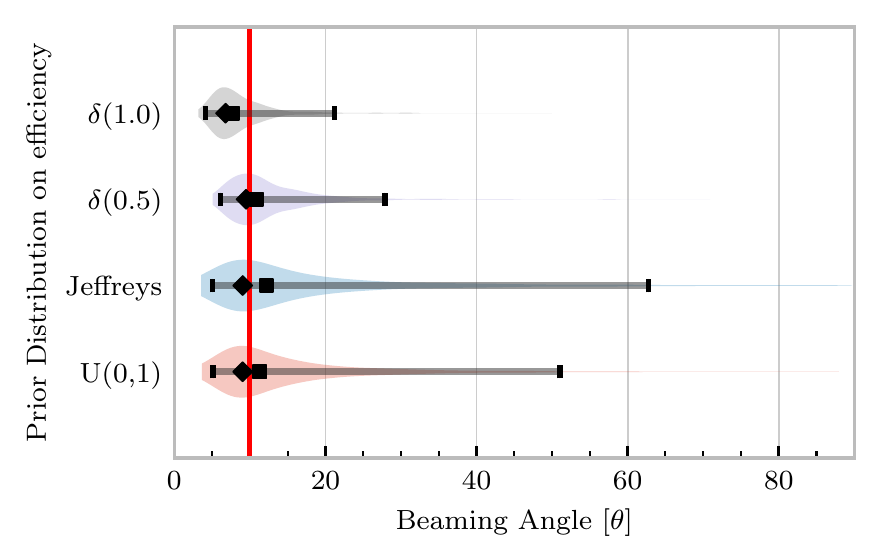}}
\caption{The procedure used to produce figure \ref{fig:injjetposterio2016} was repeated for the observing time and the horizon distance of the 2016--2017 observing scenario, with a GRB rate of $28.0 \,\yr^{-1}$ used to induce a beaming angle of $\theta \approx 10^{\circ}$.
  \label{fig:injjetposterio2017}}
\end{figure}
\begin{table}
  \centering
  \begin{tabular}{lrrrr}
    \toprule
    Prior & Lower & MAP & Median & Upper\\
          & [$^\circ$] & [$^\circ$]& [$^\circ$]& [$^\circ$] \\
    \colrule
    $\delta(1.0)$ & 4.15	 & 6.78	 & 7.62	 & 21.17	 \\
    $\delta(0.5)$ & 6.11	 & 9.50	 & 10.88	 & 27.88	 \\
    Jeffreys & 5.05	 & 9.05	 & 12.21	 & 62.72	 \\
    U(0,1) & 5.12	 & 9.05	 & 11.29	 & 51.04	 \\
    \botrule
\end{tabular}
\caption{Summary of the beaming angle posteriors from figure
  \ref{fig:injjetposterio2017}, for the 2016--2017 observing scenario,
  with an artificial GRB rate imposed to produce a target beaming
  angle of $\theta \approx 10^{\circ}$.}
  \label{tab:summaryinj2016}
\end{table}

Figures~\ref{fig:injjetposterio2016} and~\ref{fig:injjetposterio2017}
show the beaming angle posteriors which result from this analysis for
the 2015--2016 and 2016--2017 scenarios respectively for each choice
of prior distribution on the efficiency parameter.  Unsurprisingly,
the most accurate constraints arise when we already have the tightest
possible constraints on the \ac{SGRB} efficiency, $\epsilon$.  That
is, the beaming angle posterior arising from the $\delta$-function
prior on $\epsilon$ is the narrowest, yielding the shortest possible
credible interval.  It is well worth remembering, however, that had
we been incorrect regarding the value of $\epsilon$ when using the
$\delta$-function prior, the result would be significantly biased and
our inference on the beaming angle would be incorrect.  This
highlights the necessity of building a suitable representation of our
ignorance into the analysis.  Finally, we note that the results from
the uniform and Jeffreys distribution priors are broadly equivalent.

\subsubsection{Jet Angle Posteriors From Observing Scenarios}

\begin{figure}
\centering
{\includegraphics[width=\linewidth]{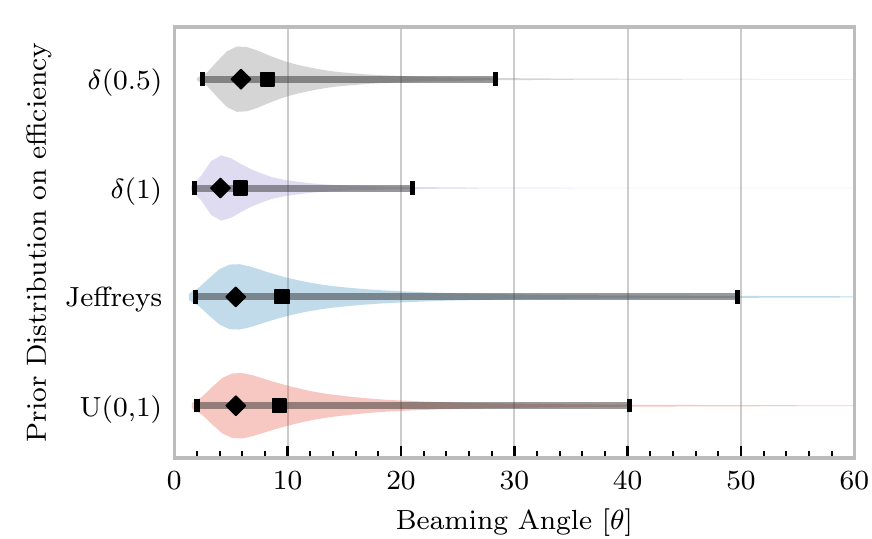}}
\caption{Beaming angle posteriors using different priors on \ac{SGRB} efficiency $\epsilon$ in the 2015--2016 observing scenario.
    \label{fig:jetposterior2016}}
\end{figure}

\begin{figure}
\centering
{\includegraphics[width=\linewidth]{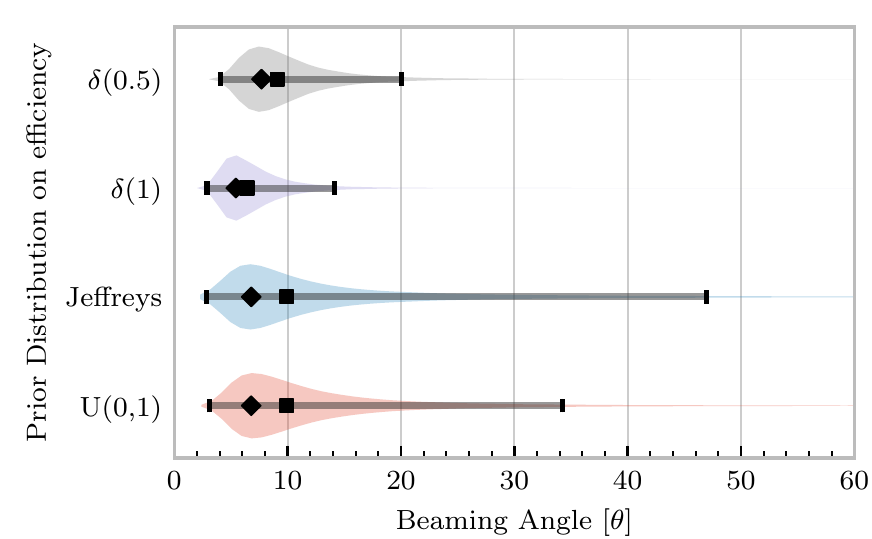}}
\caption{Beaming angle posteriors using different priors on \ac{SGRB} efficiency $\epsilon$ in the 2016--2017 observing scenario.
    \label{fig:jetposterior2017}}
\end{figure}

Figures~\ref{fig:jetposterior2016} and~\ref{fig:jetposterior2017} show
the beaming angle posteriors obtained for two of the detection
scenarios.\footnote{
    A note on implementation: rather than directly evaluating
    the beaming angle posterior in equation~\ref{eq:beam_posterior} we
    choose to sample points from the posterior using a Markov-Chain
    Monte-Carlo algorithm, implemented using the python package
    \texttt{PyMC3}~\cite{salvatier2016probabilistic}.
}\footnote{
    While we present the entire posterior for only these two
    observing scenarios in this section, we provide an overview of all
    of the observing scenarios in section~\ref{sec:future}.}
Since it is a common assumption in related literature, we also now include
a prior on the \ac{SGRB} efficiency which dictates that all \BNS produce a
\ac{SGRB}, $p(\epsilon|I)=\delta(\epsilon=1)$, as well as our previous
strong $\delta$-function prior.  For the 2016-2017 scenario where
inferences are somewhat weak (i.e., broad posteriors) due to the
sparsity of \ac{GW} detections, the uncertainties are large enough
that the results from each prior are broadly consistent.  In the 2024+
scenario, where the posterior is more peaked, it is clear that the
strong $\delta$-function priors lead to inconsistent inferences on the
\ac{SGRB} beaming angle.  The much weaker uniform and $\beta$
distributions, by contrast, are again largely consistent with each
other yielding more conservative and robust results, as well as being
a more representative expression of our state of knowledge.  The
inferences drawn from each scenario and each prior are summarised in
terms of the maximum \emph{a posteriori} measurement and the 95\%
credible interval around the maximum in
table~\ref{tab:aligo_beam_inference}.

\begin{table}
\centering
\begin{tabular}{llrrrr}
  \toprule
  Scenario & Prior & Lower & MAP & Median & Upper \\
  && [$^{\circ}$] & [$^{\circ}$]    & [$^{\circ}$]    & [$^{\circ}$]  \\
  \colrule
  2015--2016 & U(0,1) 	 & 2.00	 & 5.43	& 9.24	& 40.17	 \\
& Jeffreys 	 & 1.90	 & 5.43	& 9.50	& 49.71	 \\
& $\delta(1)$ 	 & 1.76	 & 4.07	& 5.83	& 21.04	 \\
& $\delta(0.5)$ 	 & 2.51	 & 5.88	& 8.22	& 28.35	 \\
  \colrule
  2016--2017 & U(0,1) 	 & 3.09	 & 6.78	& 9.91	& 34.23	 \\
& Jeffreys 	 & 2.85	 & 6.78	& 9.91	& 46.93	 \\
& $\delta(1)$ 	 & 2.88	 & 5.43	& 6.40	& 14.15	 \\
& $\delta(0.5)$ 	 & 4.06	 & 7.69	& 9.07	& 20.05	 \\
  \colrule
  2018--2019 & U(0,1) 	 & 6.64	 & 12.66	& 16.36	& 46.96	 \\
& Jeffreys 	 & 6.31	 & 11.76	& 15.88	& 57.48	 \\
& $\delta(1)$ 	 & 6.36	 & 9.95	& 10.97	& 18.35	 \\
& $\delta(0.5)$ 	 & 8.98	 & 14.02	& 15.55	& 26.15	 \\
  \colrule
  2020+    
& U(0,1) 	 & 8.20	 & 12.66	& 16.04	& 44.73	 \\
& Jeffreys 	 & 7.82	 & 12.21	& 15.35	& 56.99	 \\
& $\delta(1)$ 	 & 8.10	 & 10.85	& 11.12	& 14.95	 \\
& $\delta(0.5)$ 	 & 11.47	 & 14.92	& 15.75	& 21.17	 \\
  \colrule
  2024+    
& U(0,1) 	 & 9.05	 & 13.12	& 16.07	& 45.10	 \\
& Jeffreys 	 & 8.58	 & 12.21	& 15.28	& 56.30	 \\
& $\delta(1)$ 	 & 9.09	 & 11.31	& 11.30	& 14.02	 \\
           & $\delta(0.5)$ 	 & 12.82	 & 15.83	& 16.00	& 19.82	 \\
  \botrule
\end{tabular}
\caption{Summary of the beaming angle inferences for each prior in each of the observing scenarios detailed in table \ref{tab:scenarios}.
    The lower and upper values correspond to the lower and upper bounds of the 95\% Bayesian credible interval for each scenario.
    \label{tab:aligo_beam_inference}}
\end{table}

\section{Beaming Angle Constraints With No \ac{GW} Detections}
\label{sec:beaming_limits}
While GW170817 provided a situation where \ac{GW} signals from a \BNS
coalescence event \emph{were} observed, our proposed approach is also
valid in the regime where no \ac{GW} signals from \BNS coalescence
have been observed, as was true during the first observing run of the
advanced LIGO detectors when upper limits on binary merger rates were
used to place lower limits on the beaming angle~\cite{Abbott:2016ymx}.

In this scenario, our procedure is identical to before:
construct the posterior probability density function on the \BNS
coalescence rate, transform to the joint posterior on the beaming
angle and \ac{SGRB} efficiency, $\epsilon$, and marginalise over the
nuisance parameter $\epsilon$ to yield the posterior on the beaming
angle.  Now, however, rather than quoting the maximum a posteriori
estimate, together with some credible interval, we simply integrate
the beaming angle posterior from $\theta=0$ until we reach that value
which contains some desired confidence.  Thus, we obtain an upper
limit on the beaming angle, analogous to the rate upper limits set by
past LIGO observations~\cite{Colaboration:2011np}.

Figure~\ref{fig:jetposterior2016} shows the four posteriors on the
beaming angle, corresponding to the four priors on the \ac{SGRB}
efficiency, $\epsilon$, using the observing 2015--2016 observing
scenario from table~\ref{tab:scenarios}, which corresponds closely
to the conditions of the first science run of the advanced generation
of ground based \ac{GW} detectors.  We define the \emph{upper} limit
on the beaming angle as the upper limit of the 95\% credible interval
where the credible interval is defined as the narrowest interval
$(\theta^{\mathrm{ll}}, \theta^{\mathrm{ul}})$ which satisfies the expression
\begin{equation}
    \label{eq:beaming_upper_limit}
    0.95 = \int_{\theta_{\mathrm{ll}}}^{\theta^{\mathrm{ul}}} p(\theta|D,I)~\diff \theta,
\end{equation}
with $p(\theta|D,I)$ the posterior over which the interval is
computed.

Similarly we define the \emph{lower} limit as the lower
limit (2.5 percentile) of the same credible interval.  In this
non-detection scenario, we choose to compute \emph{upper} limit on the
95\% credible interval on the beaming angle.

We see that here, where the rate posterior is rather uninformative,
the results are dominated by the uncertainty in $\epsilon$: there are
substantive differences in the beaming angle upper limits yielded by
the uniform ($U(0,1)$) and $\beta$-distribution priors, while the
$\delta$-function priors yield dramatically different upper limits.
Indeed, the most stringent (and mutually incompatible) upper limits
are obtained using the strong $\delta$-function priors.  In fact,
these beaming angle upper limits are also incompatible with the values
of $3^{\circ}\mbox{-}8^{\circ}$ that have been inferred from
observations of jet breaks in \ac{SGRB}
afterglows~\cite{Fong:2013lba,2006MNRAS.367L..42P,
  2012A&A...538L...7N}.  Recall, however, from the discussion in
section~\ref{sec:sgrbs} that we interpret the beaming angle inference
from our rate measurements as the upper bound on the mean of a
population of beaming angles.  It would, therefore, seem premature to
conclude that there is tension in these results; instead, we can only
state that either the population of \acp{SGRB} have a distribution of
beaming angles with some finite width or that the fraction of \BNS
mergers which yield a \ac{SGRB} is smaller than 0.5.

It is also interesting to compare these upper limits on the beaming
angle with those in~\cite{Chen:2012qh}, where the upper limit
on the rate itself is used as a constraint (rather than transforming
the posterior).  This has the important implication that the
constraint thus obtained is the \emph{smallest} angle consistent with
the rate:
\begin{equation}
    1 - \cos \theta \geq \frac{\grbrate}{\epsilon \cbcrate^{\mathrm{ul}}},
\end{equation}
where $\cbcrate^{\mathrm{ul}}$ is the \emph{upper limit} on the \BNS
rate.  The same idea is used in~\cite{Clark:2014jpa} to estimate
beaming constraints in the advanced detector era.  Thus, when
comparing the constraints in e.g.,~\cite{Chen:2012qh} and the
upper limits obtained from the transformed posterior (i.e.,
equation~\ref{eq:beam_posterior} and figure~\ref{fig:volumevevents}), one
should remember that they are quite different quantities.  There are
two other noteworthy differences between~\cite{Chen:2012qh}
and this work: (i) the rate upper limit is computed based on the
sensitivity of the initial LIGO-Virgo network (see
e.g.,~\cite{BradyFairhurst08}), which gives $\cbcrate=4.5\times
10^{-4}$\,Mpc$^{-3}$yr$^{-1}$ (as compared with $\cbcrate=1.3 \times
10^{-4}$\,Mpc$^{-3}$yr$^{-1}$ from the analysis
in~\cite{Colaboration:2011np}); and (ii) it is implicitly assumed that
\emph{all} \BNS mergers yield an \ac{SGRB}.  That is, there is no
factor or $\epsilon$ to account for the unknown fraction of mergers
which successfully launch an \ac{SGRB} jet.  With these differences
noted, the lower bound on the beaming angle is found to be $\theta
\geq 0.8^{\circ}$, as compared with the lower limit of the 95\%
credible interval $\theta^{\mathrm{ll}}=1.76^{\circ}$ when assuming
$\epsilon=1$, and the 2015-2016 observing scenario.

\section{Beaming Angle Constraints in Future Scenarios}
\label{sec:future}
With the advent of gravitational wave astronomy, and with the
expectation of the detection of \BNS gravitational wave signals during
the lifetime of the advanced detectors it will become possible to
place further constraints on the 95\% credible interval of the SGRB
beaming angle, as both the searched 4-volume of space increases, and
the observed rate of gravitational wave \BNS events is established.

\begin{figure}
\centering
\includegraphics[width=\linewidth]{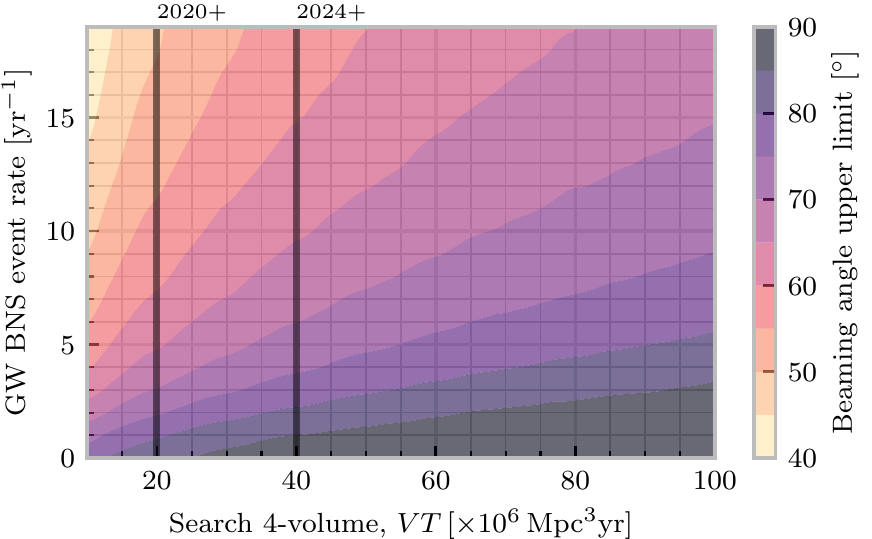}
\caption{\label{fig:volumevevents} The upper-bound of the 95\%
  credible interval on the beaming angle as a function of the rate of
  observed gravitational wave \BNS events and the observed search
  4-volume, taking a Jeffreys prior on the efficiency of \ac{GRB}
  production from \BNS{} events. The search volumes corresponding to
  observing scenarios are marked as vertical lines on the plot, with
  each line assuming that observations are carried out over the period
  of one year, achieving the search volume outlined in table
  \ref{tab:scenarios}.}
\end{figure}

\begin{figure}
\centering
\includegraphics[width=\linewidth]{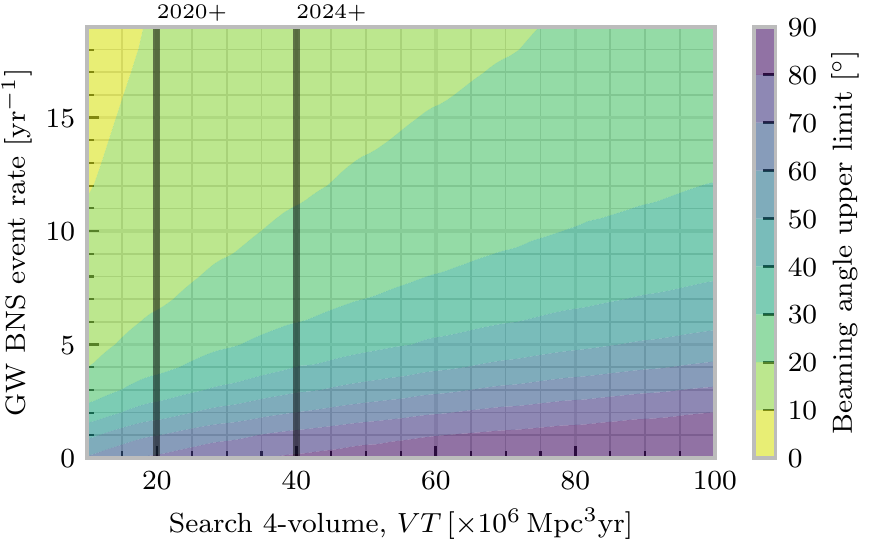}
\caption{\label{fig:volumeveventse1} The upper-bound of the 95\%
  credible interval on the beaming angle as a function of the rate of
  observed gravitational wave \BNS events and the observed search
  4-volume, assuming that all \BNS events produce a \ac{GRB}. The
  search volumes corresponding to observing scenarios are marked as
  vertical lines on the plot, with each line assuming that
  observations are carried out over the period of one year, achieving
  the search volume outlined in table \ref{tab:scenarios}.}
\end{figure}

In figure \ref{fig:volumevevents} we present the inferred upper-limit
on the 95\% credible interval for a range of search 4-volumes and
gravitational wave event rates; overlayed on this plot are indications
of the anticipated annual search volume for the advanced LIGO
detectors in each of the observing scenarios detailed in table
\ref{tab:scenarios}. These limits were determined by assuming a
Jeffreys prior on the efficiency parameter of the model, and following
the same procedure used to produce the posteriors in figures
\ref{fig:jetposterior2016} and \ref{fig:jetposterior2017}. In figure
\ref{fig:volumeveventse1} we present a similar plot, showing the upper
limits on the beaming angle under the stronger assumption that every
\BNS event also produces a \ac{GRB}.

\begin{figure}
\centering
\includegraphics[width=\linewidth]{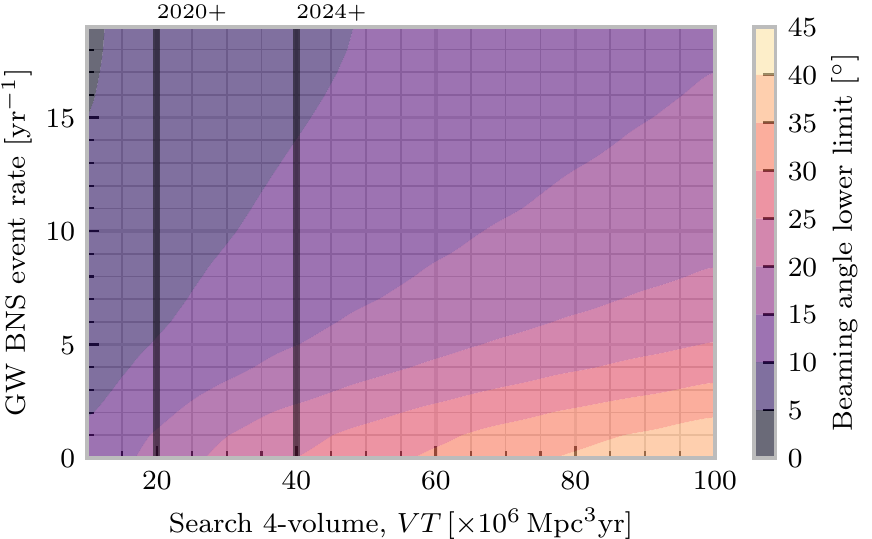}
\caption{\label{fig:volumevevents_lower} The lower-bound of the 95\%
  credible limit on the beaming angle as a function of the observed
  number of events and the observed search 4-volume, taking a Jeffreys
  prior on the efficiency of \ac{GRB} production from \BNS
  events. The search volumes corresponding to observing scenarios
  are marked as vertical lines on the plot.}
\end{figure}

\begin{figure}
\centering
\includegraphics[width=\linewidth]{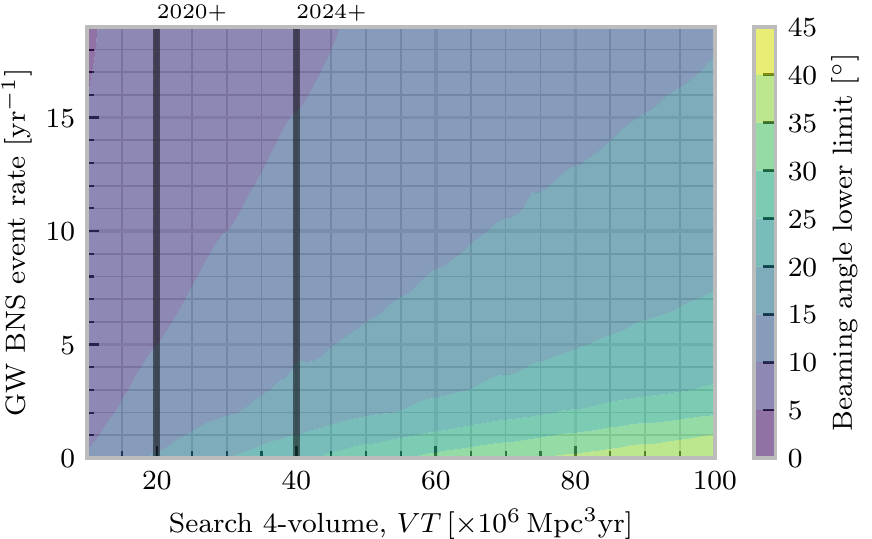}
\caption{\label{fig:volumevevents_lowere1} The lower-bound of the 95\%
  credible limit on the beaming angle as a function of the observed
  number of events and the observed search 4-volume, assuming that
  every gravitational wave \BNS event produces a \ac{GRB}. The search
  volumes corresponding to observing scenarios are marked as vertical
  lines on the plot.}
\end{figure}

The lower limit (the 2.5\% of the posterior) for the same range of
scenarios is plotted in figure \ref{fig:volumevevents_lower}, with the
same anticipated detector search volumes plotted, again assuming a
Jeffreys prior on the efficiency, and in figure
\ref{fig:volumevevents_lowere1} we present those lower limits under
the assumption that every \BNS event produces a \ac{GRB}.

\section{Conclusion}

We have presented a Bayesian analysis which demonstrates the ability
of the current generation of advanced \ac{GW} detectors to make observations
that allow for the inference of \ac{SGRB} jet beaming angles. In doing so we
have made minimal assumptions about the processes which produce the jet, other
than that \BNS mergers are the progenitors and that, other than for rare nearby
cases like GW170817, \acp{SGRB} are observed only by observers within the cone
of the jet.

We demonstrate that with a year's worth of gravitational wave
observations by the 2-detector \ac{aLIGO} network during its 2016-2017
observing run, and assuming a single \BNS detection, that we can place
a lower limit of $2.85^{\circ}$, and an upper limit of $46.93^{\circ}$
on the jet beaming angle, given an uninformative prior on the
efficiency at which \BNS events produce observable
\acp{SGRB}. Assuming that \emph{all} \BNS produce an observable
\acp{SGRB} we can narrow these limits to between $2.88^{\circ}$ and
$14.15^{\circ}$.

When the advanced LIGO design sensitivity is acheived in 2020 the
observation of 10 \BNS events in gravitational waves is sufficient to
place an upper limit of $56.99^{\circ}$ on the jet beaming angle, and can
establishing the limit on the beaming angle to be between
$7.82^{\circ}$ and $56.99^{\circ}$, assuming an uninformative prior on
the \acp{SGRB} production efficiency. These limits narrow between
$8.10^{\circ}$ and $14.95^{\circ}$ if perfect efficiency is assumed.


\acknowledgements{
The authors wish to thank Martin Hendry for many insightful and
valuable discussions during the development of this technique.

DW is supported by the Science and Technology Research Council (STFC)
grant ST/N504075/1. JC acknowledges support from NSF awards
PHYS-1505824 and PHYS-1505524SH.  ISH is supported by STFC grant
ST/L000946/1. This document has been assigned the control number
LIGO-P1600102 by the LIGO document control centre.

The python code used to produce this analysis is available as both a
Jupyter (iPython) notebook, and plain python scripts via Zonodo (doi:
10.5281/zenodo.1066019), along with the data used to produce figures
\ref{fig:volumevevents} and \ref{fig:volumevevents_lower}.}

\software{
The analysis presented in this manuscript made extensive use of the
\texttt{numpy}~\cite{numpy} and \texttt{pymc3}~\cite{pymc3} python
packages, while the figures were produced using
\texttt{matplotlib}~\cite{Hunter:2007}.}

\bibliography{grb_beams_paper}

\begin{thebibliography}{}
\expandafter\ifx\csname natexlab\endcsname\relax\def\natexlab#1{#1}\fi
\providecommand{\url}[1]{\href{#1}{#1}}

\bibitem[{Abadie {et~al.}(2012{\natexlab{a}})}]{Abadie:2012bz}
Abadie, J., {et~al.} 2012{\natexlab{a}}, Astrophys. J., 755, 2

\bibitem[{Abadie {et~al.}(2012{\natexlab{b}})}]{Colaboration:2011np}
---. 2012{\natexlab{b}}, Phys. Rev., D85, 082002

\bibitem[{Abbott {et~al.}(2008)}]{Abbott:2007rh}
Abbott, B., {et~al.} 2008, Astrophys. J., 681, 1419

\bibitem[{Abbott {et~al.}(2013)}]{Aasi:2013wya}
Abbott, B.~P., {et~al.} 2013, arXiv:1304.0670, [Living Rev. Rel.19,1(2016)]

\bibitem[{Abbott {et~al.}(2016{\natexlab{a}})}]{Abbott:2016ymx}
---. 2016{\natexlab{a}}, Astrophys. J., 832, L21

\bibitem[{Abbott {et~al.}(2016{\natexlab{b}})}]{Abbott:2016blz}
---. 2016{\natexlab{b}}, Phys. Rev. Lett., 116, 061102

\bibitem[{Abbott {et~al.}(2016{\natexlab{c}})}]{Abbott:2016nmj}
---. 2016{\natexlab{c}}, Phys. Rev. Lett., 116, 241103

\bibitem[{Abbott {et~al.}(2017{\natexlab{a}})}]{TheLIGOScientific:2017qsa}
---. 2017{\natexlab{a}}, Phys. Rev. Lett., 119, 161101

\bibitem[{Abbott {et~al.}(2017{\natexlab{b}})}]{Monitor:2017mdv}
---. 2017{\natexlab{b}}, Astrophys. J., 848, L13

\bibitem[{{Abbott} {et~al.}(2017){Abbott}, {Abbott}, {Abbott}, {Acernese},
  {Ackley}, {Adams}, {Adams}, {Addesso}, {Adhikari}, {Adya}, \&
  et~al.}]{2017ApJ...848L..12A}
{Abbott}, B.~P., {Abbott}, R., {Abbott}, T.~D., {et~al.} 2017, Astrophysical
  Journal Letters, 848, L12

\bibitem[{Abbott {et~al.}(2017)}]{Abbott:2017xzu}
Abbott, B.~P., {et~al.} 2017, Nature, arXiv:1710.05835

\bibitem[{{Allen} {et~al.}(2012){Allen}, {Anderson}, {Brady}, {Brown}, \&
  {Creighton}}]{2012PhRvD..85l2006A}
{Allen}, B., {Anderson}, W.~G., {Brady}, P.~R., {Brown}, D.~A., \& {Creighton},
  J.~D.~E. 2012, \prd, 85, 122006

\bibitem[{Blandford \& Znajek(1977)}]{Blandford:1977ds}
Blandford, R.~D., \& Znajek, R.~L. 1977, Mon. Not. Roy. Astron. Soc., 179, 433

\bibitem[{{Blinnikov} {et~al.}(1984){Blinnikov}, {Novikov}, {Perevodchikova},
  \& {Polnarev}}]{Blinnikov1984}
{Blinnikov}, S.~I., {Novikov}, I.~D., {Perevodchikova}, T.~V., \& {Polnarev},
  A.~G. 1984, SvAL, 10, 177

\bibitem[{{Brady} \& {Fairhurst}(2008)}]{BradyFairhurst08}
{Brady}, P.~R., \& {Fairhurst}, S. 2008, Classical and Quantum Gravity, 25,
  105002

\bibitem[{Bromberg {et~al.}(2013)Bromberg, Nakar, Piran, \&
  Sari}]{Bromberg:2012gp}
Bromberg, O., Nakar, E., Piran, T., \& Sari, R. 2013, Astrophys. J., 764, 179

\bibitem[{Chen \& Holz(2013)}]{Chen:2012qh}
Chen, H.-Y., \& Holz, D.~E. 2013, Phys. Rev. Lett., 111, 181101

\bibitem[{Clark {et~al.}(2015)Clark, Evans, Fairhurst, Harry, Macdonald,
  Macleod, Sutton, \& Williamson}]{Clark:2014jpa}
Clark, J., Evans, H., Fairhurst, S., {et~al.} 2015, Astrophys. J., 809, 53

\bibitem[{Dal~Canton {et~al.}(2014)}]{Canton:2014ena}
Dal~Canton, T., {et~al.} 2014, Phys. Rev., D90, 082004

\bibitem[{Dietz(2011)}]{Dietz:2010eh}
Dietz, A. 2011, Astron. Astrophys., 529, A97

\bibitem[{Eichler {et~al.}(1989)Eichler, Livio, Piran, \&
  Schramm}]{Eichler:1989ve}
Eichler, D., Livio, M., Piran, T., \& Schramm, D.~N. 1989, Nature, 340, 126

\bibitem[{Fong {et~al.}(2014)Fong, Berger, Metzger, Margutti, Chornock,
  {et~al.}}]{Fong:2013lba}
Fong, W., Berger, E., Metzger, B.~D., {et~al.} 2014, Astrophys. J., 780, 118

\bibitem[{Fong {et~al.}(2017)}]{Fong:2017ekk}
Fong, W., {et~al.} 2017, Astrophys. J., 848, L23

\bibitem[{Fong {et~al.}(2015)Fong, Berger, Margutti, \&
  Zauderer}]{Fong:2015oha}
Fong, W.-f., Berger, E., Margutti, R., \& Zauderer, B.~A. 2015, Astrophys. J.,
  815, 102

\bibitem[{Galama {et~al.}(1998)}]{Galama:1998ea}
Galama, T.~J., {et~al.} 1998, Nature, 395, 670

\bibitem[{Giacomazzo {et~al.}(2013)Giacomazzo, Perna, Rezzolla, Troja, \&
  Lazzati}]{Giacomazzo:2012zt}
Giacomazzo, B., Perna, R., Rezzolla, L., Troja, E., \& Lazzati, D. 2013,
  Astrophys. J., 762, L18

\bibitem[{Goldstein {et~al.}(2017)}]{Goldstein:2017mmi}
Goldstein, A., {et~al.} 2017, Astrophys. J., 848, L14

\bibitem[{Gottlieb {et~al.}(2017)Gottlieb, Nakar, Piran, \&
  Hotokezaka}]{Gottlieb:2017pju}
Gottlieb, O., Nakar, E., Piran, T., \& Hotokezaka, K. 2017, arXiv:1710.05896

\bibitem[{{Gregory}(2010)}]{2010blda.book.....G}
{Gregory}, P. 2010, {Bayesian Logical Data Analysis for the Physical Sciences}

\bibitem[{Haggard {et~al.}(2017)Haggard, Nynka, Ruan, Kalogera, Bradley~Cenko,
  Evans, \& Kennea}]{Haggard:2017qne}
Haggard, D., Nynka, M., Ruan, J.~J., {et~al.} 2017, Astrophys. J., 848, L25

\bibitem[{Hunter(2007)}]{Hunter:2007}
Hunter, J.~D. 2007, Computing In Science \& Engineering, 9, 90

\bibitem[{Kasliwal {et~al.}(2017)}]{Kasliwal:2017ngb}
Kasliwal, M.~M., {et~al.} 2017, Science, arXiv:1710.05436

\bibitem[{Kouveliotou {et~al.}(1993)Kouveliotou, Meegan, Fishman, Bhyat,
  Briggs, {et~al.}}]{Kouveliotou:1993yx}
Kouveliotou, C., Meegan, C.~A., Fishman, G.~J., {et~al.} 1993, Astrophys. J.,
  413, L101

\bibitem[{Lee \& Ramirez-Ruiz(2007)}]{Lee:2007js}
Lee, W.~H., \& Ramirez-Ruiz, E. 2007, New J. Phys., 9, 17

\bibitem[{{LIGO Scientific Collaboration} \& {Virgo
  Collaboration}(2010)}]{rates_paper}
{LIGO Scientific Collaboration}, \& {Virgo Collaboration}. 2010, Classical and
  Quantum Gravity, 27, 173001.
\newblock \url{http://stacks.iop.org/0264-9381/27/i=17/a=173001}

\bibitem[{MacFadyen \& Woosley(1999)}]{MacFadyen:1998vz}
MacFadyen, A., \& Woosley, S.~E. 1999, Astrophys. J., 524, 262

\bibitem[{Nakar(2007)}]{Nakar:2007yr}
Nakar, E. 2007, Phys. Rept., 442, 166

\bibitem[{Narayan {et~al.}(1992)Narayan, Paczynski, \& Piran}]{Narayan:1992iy}
Narayan, R., Paczynski, B., \& Piran, T. 1992, Astrophys. J., 395, L83

\bibitem[{{Nicuesa Guelbenzu} {et~al.}(2012){Nicuesa Guelbenzu}, {Klose},
  {Kr{\"u}hler}, {Greiner}, {Rossi}, {Kann}, {Olivares}, {Rau}, {Afonso},
  {Elliott}, {Filgas}, {K{\"u}pc{\"u} Yolda{\c s}}, {McBreen}, {Nardini},
  {Schady}, {Schmidl}, {Sudilovsky}, {Updike}, \& {Yolda{\c
  s}}}]{2012A&A...538L...7N}
{Nicuesa Guelbenzu}, A., {Klose}, S., {Kr{\"u}hler}, T., {et~al.} 2012, \aap,
  538, L7

\bibitem[{Nissanke {et~al.}(2010)Nissanke, Holz, Hughes, Dalal, \&
  Sievers}]{Nissanke:2009kt}
Nissanke, S., Holz, D.~E., Hughes, S.~A., Dalal, N., \& Sievers, J.~L. 2010,
  Astrophys. J., 725, 496

\bibitem[{Nitz {et~al.}(2017)Nitz, Harry, Brown, Biwer, Willis, Canton,
  Pekowsky, Dent, Williamson, Capano, De, Cabero, Machenschalk, Kumar,
  Massinger, Lenon, Fairhurst, Reyes, Nielsen, shasvath, Pannarale, Singer,
  Macleod, Babak, Gabbard, Veitch, Sugar, Zertuche, Couvares, \&
  Bockelman}]{alex_nitz_2017_844934}
Nitz, A., Harry, I., Brown, D., {et~al.} 2017, ligo-cbc/pycbc: O2 Production
  Release 17, , , doi:10.5281/zenodo.844934.
\newblock \url{https://doi.org/10.5281/zenodo.844934}

\bibitem[{Paczy\'{n}ski(1991)}]{Paczynski:1991aq}
Paczy\'{n}ski, B. 1991, Acta Astron., 41, 257

\bibitem[{{Panaitescu}(2006)}]{2006MNRAS.367L..42P}
{Panaitescu}, A. 2006, \mnras, 367, L42

\bibitem[{Pannarale \& Ohme(2014)}]{Pannarale:2014rea}
Pannarale, F., \& Ohme, F. 2014, Astrophys. J., 791, L7

\bibitem[{Rosswog \& Ramirez-Ruiz(2002)}]{Rosswog:2002rt}
Rosswog, S., \& Ramirez-Ruiz, E. 2002, Mon. Not. Roy. Astron. Soc., 336, L7

\bibitem[{Salvatier {et~al.}(2016{\natexlab{a}})Salvatier, Wiecki, \&
  Fonnesbeck}]{salvatier2016probabilistic}
Salvatier, J., Wiecki, T.~V., \& Fonnesbeck, C. 2016{\natexlab{a}}, PeerJ
  Computer Science, 2, e55

\bibitem[{Salvatier {et~al.}(2016{\natexlab{b}})Salvatier, Wiecki, \&
  Fonnesbeck}]{pymc3}
---. 2016{\natexlab{b}}, PeerJ Computer Science, 2, e55.
\newblock
  \url{http://dblp.uni-trier.de/db/journals/peerj-cs/peerj-cs2.html#SalvatierWF16}

\bibitem[{Savchenko {et~al.}(2017)}]{Savchenko:2017ffs}
Savchenko, V., {et~al.} 2017, Astrophys. J., 848, L15

\bibitem[{Schutz(1986)}]{Schutz:1986gp}
Schutz, B.~F. 1986, Nature, 323, 310

\bibitem[{Usman {et~al.}(2016)}]{Usman:2015kfa}
Usman, S.~A., {et~al.} 2016, Class. Quant. Grav., 33, 215004

\bibitem[{van~der Walt {et~al.}(2011)van~der Walt, Colbert, \&
  Varoquaux}]{numpy}
van~der Walt, S., Colbert, S.~C., \& Varoquaux, G. 2011, Computing in Science
  \& Engineering, 13, 22.
\newblock \url{http://aip.scitation.org/doi/abs/10.1109/MCSE.2011.37}

\bibitem[{Woosley \& Bloom(2006)}]{Woosley:2006fn}
Woosley, S.~E., \& Bloom, J.~S. 2006, Ann. Rev. Astron. Astrophys., 44, 507

\bibitem[{Zhang {et~al.}(2009)}]{Zhang:2009uf}
Zhang, B., {et~al.} 2009, Astrophys. J., 703, 1696

\end{thebibliography}

\end{document}